

\def\etal{et al.~}

%
\input user$work:[tex.inputs]aa.mtm
%
%
\MAINTITLE={HST Observations of Globular Clusters in M31.
 I: Surface Photometry of 13 Objects $^*$ }
\AUTHOR={F. Fusi Pecci@1, P. Battistini@2, O. Bendinelli@2,
F. B\`onoli@2, C. Cacciari@1, S. Djorgovski@3, L. Federici@1,
F.R. Ferraro@1, G. Parmeggiani@1, N. Weir@3, and F. Zavatti@2}
\SENDOFF={ F. Fusi Pecci }
\INSTITUTE={@1 Osservatorio Astronomico, Via Zamboni 33, 40126 Bologna,
                Italy
	    @2 Dipartimento di Astronomia, Universit\`a di Bologna, CP 596,
 	            40100 Bologna, Italy
	    @3 Division of Physics, Mathematics and Astronomy, Caltech,
		Pasadena, CA 91125, USA.}
\RECDATE={ ????? }
\ACCDATE={ ????? }

%

%
\SUMMARY={ We present the initial results of a study of globular clusters
in M31, using the Faint Object Camera (FOC) on the Hubble Space
Telescope (HST).  The sample
of objects consists of 13 clusters spanning a range of properties.  Three
independent image deconvolution techniques were used in order to compensate for
the optical problems of the HST, leading to mutually fully consistent results.
We present detailed tests and comparisons to determine the reliability and
limits of these deconvolution methods, and conclude that high-quality surface
photometry of M31 globulars is possible with the HST data.  Surface brightness
profiles have been extracted, and core radii, half-light radii, and central
surface brightness values have been measured for all of the clusters in the
sample. Their comparison with the values derived from ground-based observations
indicates the later to be systematically and strongly biased by the seeing
effects, as it may be expected.  A comparison of the structural parameters with
those of the Galactic globulars shows that the structural properties of the M31
globulars are very similar to those of their Galactic counterparts.  A
candidate for a post-core-collapse cluster, Bo 343 = G 105, has been already
identified from these data; this is the first such detection in the M31
globular cluster system. }
%
%
\KEYWORDS={ Galaxies: individual -- Galaxies: M31 --
Clusters: globular -- survey: HST}
%
%
\THESAURUS={ 04.19.1; 11.09.1 M31; 11.19.4; 03.20.1 }
%
%
\maketitle
%
%
%
\vfootnote{$^*$} {Based on observations
with the NASA/ESA Hubble Space Telescope, obtained at the Space Telescope
Science Institute, which is operated by the Association of Universities for
Research in Astronomy, Inc., under NASA contract NAS5-26555.}
\titlea{Introduction}
Studies of globular clusters are important for many diverse fields of
astronomy.  Globulars represent effective laboratories to study stellar
evolution and stellar dynamics.  Globular cluster systems can be used to
map and understand the properties and evolution of their parent galaxies
(cf. Harris 1991 for a recent review).  They can also form templates of
stellar
populations useful for studies and modeling of spectroscopic evolution of
galaxies.  The best constraints we now have for the age of the universe also
come from age-dating of globular clusters.  So far, only the globular cluster
system of our own Galaxy has been studied in depth, with the systems of the
Magellanic Clouds' clusters coming second.

The closest globular cluster system beyond the halo of our Galaxy is that of
M31, and its studies are a natural next step.  It represents the largest sample
of globular clusters found in the Local Group and it is sufficiently nearby to
allow a variety of ground-based observations. Numerous studies to date have
resulted in a considerable data-base (for reviews and references see, e.g.,
Battistini \etal 1993a,b,  Federici \etal 1993, Brodie 1993,
 Cohen 1993, Fusi Pecci \etal 1993a, Huchra 1993, Racine 1993).
  Like that of our Galaxy, and unlike the clusters of the
Magellanic Clouds, it is an old, evolved system.  It thus represents a good
test case to look for both the similarities with, and possible differences from
the Galactic globular cluster system.

However, ground-based studies of M31 globulars are severely seeing-limited. At
the M31 distance ($\sim $ 700 kpc) a linear size of 1 pc corresponds to about
0.3 arcsec, so that only the brightest stars in the outskirts of globular
clusters can be resolved and reliably measured from the ground even in
exceptional seeing conditions (cf. Heasley \etal 1988, Christian\&Heasley
1991, Racine\&Harris 1991, Racine 1993).  Observations with the Hubble Space
Telescope (HST) can largely overcome this difficulty, and may lead to a
fundamental improvement in our knowledge and understanding of M31 globulars.

With this in mind, we initiated an imaging and photometric study of a selected
sample of M31 globular clusters, using the HST.  Here we present the initial
results of our long-term program.  Our goals include measurements of surface
brightness profiles and color-magnitude diagrams (CMDs) down to the
horizontal branch (HB) level.  We intend to use these measurements to
address the following issues and questions:

\noindent
\item{(a)} The HB magnitude level and morphology, and its dependence on
metallicity. This, in turn, is relevant for the distance-scale problem.

\noindent
\item{(b)} The stellar populations of M31 globulars and the comparison with
their Galactic counterparts.  This is also relevant for the use of globulars
as templates for population synthesis studies in galaxies.

\noindent
\item{(c)} The structural and dynamical parameters of M31 globulars, their
possible dependence on the position within M31, and comparisons with those of
the Galactic system.  Any radial trends found may be used to probe the
dynamical properties of M31 itself.

\smallskip\noindent
The problems with the optical performance of the HST present severe limits in
our ability to address the goals (a) and (b).  Thus, we were granted observing
time (through the HST Cycle 1 GO program \#2583: {\it Surface Photometry of a
Sample of Globular Clusters in M31}) only for the purpose of addressing the
goal (c) of our entire scientific program.  So far, 13 clusters have been
observed to complete the observations planned for the HST Cycle 1.

Some of our preliminary results from the early data reductions, and various
tests and simulations, including the surface photometry, image deconvolutions,
and photometry of individual stars, have been presented elsewhere (Bendinelli
\etal 1992; Cacciari \etal 1992).  We have also presented the first detection
of a central cusp (presumably a collapsed core) in a globular cluster in M31,
G 105 = Bo 343 (Bendinelli \etal 1993).  Here we give a complete report on the
observations carried out so far, a description of the basic data analysis, and
a presentation and discussion of the main results.  Section 2 describes the
sample selection and the observations.  Section 3 describes the results of the
surface photometry on both the observed and deconvolved image, and discusses
and compares the three different deconvolution methods we use.  In Section 4 we
present the derived structural parameters of our sample clusters, their
comparison with previous results obtained from the ground for M31 globulars,
and a comparison with the properties of Galactic globulars.  Finally, in
Section
5 we present some concluding remarks and discuss the future prospects.

\titlea{The cluster sample and observations}

\titleb{The observed sample}
We selected the sample of clusters for this study as follows.  We started with
a joint catalog of M31 globular clusters, from a merged list of the three basic
lists in the literature: Sargent \etal (1977) (commonly designated
as G-number),
Crampton \etal (1985), and Battistini \etal (1980, 1987, 1993a,b), (commonly
designated with Bo-number).  Many objects in these catalogs overlap, e.g.,
G 105 = Bo 343, etc.  From this joint catalog we then selected clusters
satisfying as many of the following criteria as possible:

\smallskip\noindent
(1) Objects confirmed as globular clusters from high-resolution ground-based
imaging or spectroscopy, and well separated from any nearby objects.

\smallskip\noindent
(2) Bright, with V$>$17 (M$_V \sim -$7.5).

\smallskip\noindent
(3) Sufficiently unreddened and projected on a low background.

\smallskip\noindent
(4) Clusters with reasonable metallicity estimates, spanning a large
metallicity range.

\smallskip\noindent
(5) Clusters at a range of galactocentric distances, ranging from a few objects
located in the very central regions of M31, out to some of the farthest
globulars detected in M31 so far (at $\sim $30kpc).

\smallskip\noindent
(6) Clusters with at least preliminary estimates of structural parameters
(e.g., W$_{1/4}$; cf. Battistini \etal 1982, 1987, Buonanno \etal 1982 or
Crampton \etal 1985), spanning a range of concentrations.

\smallskip\noindent
(7) Clusters identified with known X-ray sources (van Speybroeck \etal 1979;
Battistini \etal 1982; Long \etal 1983; Crampton \etal 1984; Collura \etal
1990; Trinchieri\&Fabbiano 1991).

\smallskip\noindent
(8) If possible, clusters falling within the reach of HST as parallel targets,
where the primary targets were determined by some of the criteria (1) -- (7).

\smallskip\noindent
At the time of the first submission of our HST proposal, our catalog of the
selected primary target clusters included about 30 objects.  This number has
since increased to more than 50 objects, thanks to the subsequent observations.

We have initially included in a GTO program (\#1283) what we judged are
among the best target clusters in order to address as soon as possible our
three principal scientific goals, as described in the previous section.
However, after the GTO and GO program reassessment prompted by the HST optical
aberration, our initial full GTO and GO programs have been delayed, due to a
limited ability to construct CMDs.  Our final HST Cycle 1 GO program \#2583 was
formulated as a subset of the original scientific program, and has been
allocated
observing time (using just one filter, see below) in order to do a surface
brightness study only.  As a consequence, the clusters included in the
GTO program were delayed and the group of clusters included in the
present sample has been optimised for that goal alone.  These are not all our
first-choice objects, and included here are some clusters which we originally
ranked as being of a lower priority.  Table 1 gives the list of clusters we
actually observed with the HST in Cycle 1, and whose reductions and analysis is
presented here.  The table also includes a set of the relevant data on these
clusters, taken from the literature.
\begtab 2 cm
\tabcap{1}{List of observables for the M31 HST-program clusters.}
\endtab

\titleb{The HST observations}
All of the observations reported here were taken with the HST Faint Object
Camera
(FOC) in its $f/96$ mode, through the F430W ($B$ band) filter.  This gives
an 11 arcsec field of view, with 0.022 arcsec pixels.  For most
clusters two separate integrations were made, taken sequentially with nearly
the same (or very similar) exposure times.  We also obtained individual
exposures of isolated bright stars to serve for the measurements of the point
spread function (PSF).  The exposure times have been calculated so as to obtain
at least S/N = 5 at a radial distance of 3 arcsec, on the assumption that the
surface brightness profiles can be described by King models.  Table 2 gives
a complete log of the observations, including the maximum and average
counts per pixel in the preprocessed images.
\begtab 2 cm
\tabcap{2}{Log of HST-FOC observations (identifications from Battistini
\etal 1987 or Sargent \etal 1977).}
\endtab
We obtained repeated observations of bright stars interleaved with our globular
cluster exposures, in order to measure the PSF with the same HST focus,
tracking, and other observational setup parameters.  As it turns out, only two
out of three of these PSF exposures were useful:  in one of them the target
star was not properly centered in the frame, and the PSF wings could not be
measured adequately.

Exposures of a few of our clusters were also slightly offset with respect to
the FOC frame centers.  We measured the accurate coordinates of our clusters
using GASP frames at the ST-ECF; these should be better than the original
positions from the Battistini \etal (1987) catalog, with accuracies better than
1 or 2 arcsec.  However, it was sometimes difficult to find the adequate
guide stars.  All of our exposures were made in the standard
``fine lock'' tracking mode.  We have not detected any obvious smearing or
jitter problems in our data.

Figure 1 shows a typical example of our cluster images, G 322 = Bo 386,
and one of the PSF stars.  We can see that the clusters are clearly resolved,
but just barely.

\begfig 2 cm
\figure{1} {Row HST images of the PSF star GD 248 ({\it top left:}
intensity, {\it top right:} Log intensity), and of the M31 globular cluster
G 322 = Bo 386 ({\it bottom left:} intensity, {\it bottom right :} Log
intensity).}
\endfig

\titlea{ Data reductions and analysis}

The images have been preprocessed at STScI using standard procedures, in order
to remove the instrumental signatures, including the geometrical distortion
and flatfield corrections.  We have then
windowed the images to exclude the distorted borders which result from the
geometrical correction.  Finally, separate exposures for each cluster were
shifted into coincidence and coadded.  For some of our deconvolution tests,
we treated the individual exposures separately, in order to provide an
internal estimate of the photometric and deconvolution uncertainties.

Given the substantial amount of the total energy in the PSF wings, and the
barely-resolved nature of these cluster images, it was necessary to apply some
image deconvolution before the cluster parameters could be measured with any
degree of certainty.  We used three very different methods for this task,
namely:

(a) Regularized Multi-Gaussian technique (Bendinelli 1991; hereafter RMG).  In
this method, circularly averaged surface brightness profiles are extracted
first, and deconvolution is performed in the profile domain.

(b) Richardson-Lucy iterative technique (Richardson 1972; Lucy 1974, 1992;
hereafter RL).  In this method, the entire 2-dim. image is deconvolved, and
the surface brightness profile is extracted from the resulting image.

(c) Maximum Entropy method (Weir 1991, and references therein; hereafter MEM).
Here also images are deconvolved first, and surface brightness profiles are
extracted thereafter.

Typically we used 512 $\times$ 512 subimages for the RL deconvolutions, and
256 $\times$ 256 subimages for the MEM deconvolutions.  Larger size images were
computationally too expensive for the MEM, and most of the relevant signal was
contained within the 256 $\times$ 256 subimages anyway.  We illustrate the
results of the RL and MEM deconvolutions in Figure 2, on an example of the
cluster G 322 = Bo 386.

\begfig 2 cm
\figure{2} {Example of observed and deconvolved images of the M31  globular
cluster Bo 386 = G 322. {\it Top left:} Observed image; {\it Top right:}
MEM-deconvolved image; {\it Bottom left:} RL-deconvolved image.}
\endfig
We extracted surface brightness profiles from our images using a
Berkeley-Caltech-Bologna version of the VISTA package (originally developed at
Lick Observatory by Drs. T. Lauer and R. Stover).  Two independent methods were
used: one specifically designed for the surface photometry of (Galactic)
globular clusters, whereby circularly averaged surface brightness profiles are
extracted (for a description of methodology, see Djorgovski 1988); and one
designed for surface photometry of galaxies, in which elliptically averaged
surface brightness profiles are extracted (for a description of methodology,
see Djorgovski 1985, and references therein).  The assumption of a circular
symmetry is a good approximation in this case, and the two methods
give results in excellent mutual agreement.

We note that the HST PSF overfills slightly the FOC $f/96$ frames (11$\times$11
arcsec), and so our measurements are necessarily unreliable at radii exceeding
a few arcsec, or about 10 pc at the distance to M31.  This affects both the
surface brightness measurements through the sky subtraction uncertainties, and
the deconvolutions.  The problems are made slightly worse in the cases where
clusters were slightly offset from the frame centers.  As a result, we cannot
measure the tidal radii of our clusters, and in some cases even the half-light
radii may be affected.  However, measurements of the core parameters should be
relatively insensitive to this problem.

In addition to the surface photometry and measurements of structural parameters
of our clusters, we also attempted a stellar photometry for some of them.  We
used an updated version of the ROMAFOT package for this purpose (Buonanno
\etal 1983, Buonanno\&Iannicola 1988).  This is a far more demanding
task than the surface photometry, and much more work is required before any
reliable results can be presented.  We thus postpone the discussion of the
stellar photometry for a future paper.  A preliminary report and a discussion
of some of the problems encountered have been presented by
Cacciari \etal (1992) and Bendinelli \etal (1992).

We now discuss our profile and image deconvolutions in some detail.  We note
that all deconvolutions are driven by the quality of the input data: the
S/N ratio, and the adequacy of the PSF assumed.

\titleb{Profile deconvolutions using the RMG method}

The description of the Regularized Multi-Gaussian (RMG) deconvolution method,
as applicable to extended objects and PSFs with a (nearly) circular symmetry,
has been presented by Bendinelli (1991).  Briefly, the surface brightness
profiles of both the object to be deconvolved and the PSF are represented as
series of Gaussians, the number of terms being determined by the requirement
that the model is a good fit to the data (usually less than 10 terms are
needed).  Figure 3 illustrates a multi-Gaussian representation of the observed
PSF profiles, and the residuals from the fit, whose parameters,
identical for the two stars, are reported in Table 3. The resulting
deconvolved clusters profiles are then also expressed as a series of Gaussians,
described by the parameters listed in Table 4.

\begfig 2 cm
\figure{3}{Regularized Multi-Gaussian deconvolution of the two stars
GD248 and G158-100. {\it Squares:}
observed values; {\it lines:} adopted FOC f/96 PSF brightness profiles
described by {\it the same} Multi Gaussians fit, used in RMG deconvolutions.
In the bottom boxes the residuals after fit are shown.}
\endfig
\begtab 2 cm
\tabcap{3} {Parameters of the four Gaussian components for the PSF
Multi-Gaussian expansion.}
\endtab
\begtab 2 cm
\tabcap{4}{Parameters of the eight Gaussian components for the
Multi-Gaussian expansion of the deconvolved M31 cluster profiles.}
\endtab
We note that the circular or elliptical symmetry assumptions allow this method
to deal with one-dimensional surface brightness profiles directly.  This yields
computational convenience and speed, as well as numerical stability.
However, the flip side is that a fully deconvolved two-dimensional image is not
produced, which may or may not be a drawback, depending on the particular
scientific application.

The first direct application of this method to ground-based observations of
globular clusters in M31, along with extensive reliability tests and
comparisons, has been reported by Bendinelli \etal (1990). An extension of the
method to images and PSFs with elliptical symmetry has been presented by
Monnet \etal (1992).  Zavatti \etal (1991) have explored the applicability of
the RMG method to deconvolutions of HST images of extended objects.

Figure 4 presents a direct comparison of the observed and RMG-deconvolved
profiles (except Bo 343, see below). In Figure 5, the latter ones are
compared with those obtained from the other two methods.  To verify further
the reliability of the adopted procedure, we have also performed extensive
tests of the method using simulated data.  Some of these
tests, as well as the comparisons with the RL and MEM techniques, have been
presented elsewhere by Bendinelli \etal (1992, 1993).
\begfig 2 cm
\figure{4}{Observed {\it (squares)} and RMG-deconvolved {\it (lines)}
profiles of the individual clusters.}
\endfig
\titleb {Image deconvolutions using the RL method}
The RL method is now widely employed for the deconvolutions of HST images, due
to its relatively high speed and simplicity.
We have used two implementations of the
method: the procedure provided in the IRAF package (1991 release), and a new,
``accelerated'' version (Lucy 1992).
\begfig 2 cm
\figure{5}{Comparison of the RMG, RL, and MEM deconvolutions of the
individual clusters. {\it Line}: RMG-deconvolved profile;
{\it open squares}: MEM; {\it filled squares}: RL. The profiles have been
vertically shifted by $\Delta Log = 1$ for the sake of clarity.}
\endfig

With the IRAF version, we first made tests using 30, 60, 90, or 120 iterations,
always assuming the noise level of 0.5, and the limiting $\chi^2$ of 0.001. For
less than 60 iterations, we found prominent geometrical distortions.  Only
negligible changes appeared with more than 60 iterations.  We then used the new
``accelerated'' version with 10, 30, 50, or 70 iterations,
and the same parameters as given above.  We decided that 60 iterations in the
new version is the optimal number to apply to RL deconvolutions of all our
objects.  We note that 1 iteration in the ``accelerated'' version corresponds
to about 5 in the IRAF implementation.

Surface brightness profiles have been extracted from the deconvolved images.
The results of the RL deconvolution are shown in Figure 5, where they can
be compared to those obtained from the other two methods.
\titleb{Image deconvolutions using the MEM}
The principal advantage of the Maximum Entropy Method is that it can be
rigorously shown that it is the statistically optimal deconvolution technique,
if applied correctly (cf. Weir\&Djorgovski 1990, Weir 1991, and references
therein). The principal drawback is its great computational cost.  We used the
implementation of MEM described by Weir (1991), which uses the Gull-Skilling
MemSys5 library.  Some applications of this package to the real and simulated
HST data have also been described by Weir\&Djorgovski (1990) and King \etal
(1991), where more details and references can be found.

There are two important new features of this implementation, which are great
improvements over the classical (pre-1990) MEM: first, the use of multi-channel
restorations, with different intrinsic correlation functions (ICFs) in each;
and second, the ability to restore multiple exposures of the same object
simultaneously.  The first feature improves the restoration of extended
structures where a range of spatial scales is present.  The second one provides
better constraints on the restored images, using all of the available
information simultaneously.

Surface brightness profiles have been extracted from the deconvolved images
and the results are shown in Figure 5.
Historically, MEM was often used to sharpen and resolve images, but it has
never been used well in astronomical {\it surface photometry}.  Perhaps the
main reason for this is a bias inherent to the method:  MEM tends to suppress
faint objects or features near the bright ones; in this context, faint
envelopes around bright cores.  In other words, MEM tends to oversharpen the
bright cores of extended sources at the expense of their lower surface
brightness envelopes. Our experience has shown that this effect is shared
also by the RL-procedure, although to a lesser degree.

On the other hand, we have verified that the use of multiple channels
and a range of ICFs tends to smooth the sharp features and therefore minimizes
this effect, producing surface brightness profiles which are essentially
identical to those obtained via the RL restoration procedure.
Additional future tests with simulated images should determine the nature
of bias in MEM surface photometry, if there is indeed any.
\titleb{Comparisons of different methods}
Probably the safest way to estimate the uncertainties in the deconvolved images
and profiles is to compare the results obtained using different techniques.  We
present such comparisons in Figures 5 and 6.  In particular, the plots
presented in Figure 6 report the radial behaviour of the residuals of
the profiles.
We also note that all deconvolutions were computed independently and normalized
as described in  Sect. 3.6. Moreover, the brightness profiles
were derived by using two different VISTA routines.
\begfig 2 cm
\figure{6} {The radial distribution of the residuals of
the deconvolved profiles after  normalization and calibration (see Sect. 3.6).
{\it Open squares:} MEM $-$ RMG; {\it filled squares:} RL $-$ RMG.
The divergence of the MEM- and RL-deconvolved profiles
with respect to the RMG-deconvolutions in the external part of the profiles
is the {\it edge effect} discussed in Sect. 3.6.3.
To avoid any systematic bias, the profile normalization has thus been
extended for all clusters only as far as $r_{lim} = 1.5$ arcsec.}
\endfig
On the whole, the agreement between different methods is remarkably good for
all observed clusters over most of the radial range covered by the data.
This is a very encouraging sign, and an evidence that
the deconvolution results are reliable, in the sense that no obvious artifacts
are introduced by the deconvolution processes. Within this frame, it is
interesting to note that in the cluster G11 = Bo 293 a bright
(possibly Galactic foreground) star is present near the projected
cluster center. Early deconvolutions carried out without subtracting the
star contribution have led to discrepant results, while
after subtraction totally consistent profiles and structural parameters
have been obtained.
\par
We must note, however, that the fact that different deconvolutions of the same
data are mutually consistent by itself does not guarantee that the results are
also $right$.  We performed independent MEM deconvolutions of separate
exposures of a given cluster, and found the surface photometry results to be in
an excellent agreement; this suggests that the S/N is not a limiting factor
here.  However, we still use the same PSFs, and any imperfections in the PSF
could, in principle, affect the results.  We have no control over such possible
problems, but we also have no reasons to suspect that there $are$ any problems.

We further illustrate the robustness of the results in Figure 2, which shows a
comparison of deconvolved images of the cluster G 322 = Bo 386.
It is interesting to note how similar are also the structures seen at small
angular scales, obtained from the RL and MEM deconvolutions.
This, again, should be seen as a sign of consistency, but not necessarily as
an evidence for resolved stars: most of the apparent ``stars'' seen in MEM
deconvolutions do $not$ repeat between the two independent exposures of a
given cluster.  These are simply chance fluctuations in the signal, possibly
caused by an imperfect flatfielding.  This surely $is$ a S/N problem, and its
limits on the stellar photometry in our images remain to be explored in a
future paper.
\titleb{Bo 343 = G 105, a cluster with a central cusp}
In the course of this work, we have made the first detection of a
collapsed-core
globular cluster in M31, G105 = Bo 343.  More details can be found in the
paper by Bendinelli \etal (1993), and here we summarize the salient
points.

The problem of core collapse is one of the central topics in stellar dynamics
and astrophysics of globular star clusters (cf., e.g., Elson, Hut\&Inagaki
1987, and references therein).  The collapse and a subsequent recovery lead to
the appearance of a characteristic surface brightness profile morphology near
the cluster center: a power-law cusp with a slope $\sim -$0.5 to $-$1,
extending up to a few parsec out in radius.  Approximately one-fifth of all
known globulars in our Galaxy show this morphology (Djorgovski\&King 1984,
1986), which distinguishes them from the standard,
King-model clusters with flat, extended cores and steeper envelopes (King
1966).  Dynamical evolution of clusters can be speeded up by the tidal shocks
along their orbits, and a census of post-core-collapse clusters as a function
of their position in the parent galaxy can thus serve as a powerful probe of
the evolution of the system; so far, this was done only for our own Galaxy
(Djorgovski 1988; Chernoff\&Djorgovski 1989).  Core collapse can also affect
the stellar populations in significant ways which are not yet fully understood
(Djorgovski \etal 1991; Fusi Pecci \etal 1993b;
Djorgovski\&Piotto 1992, 1993).  Both of these types of effects may be
relevant for our study of the M31 globular cluster system.

The only other galaxies for which a systematic search for post-core-collapse
clusters has been done are the Magellanic Clouds (Meylan\&Djorgovski 1987).
So far, measurements of the morphology of globulars in other galaxies,
including M31, have been precluded by their distance and a seeing-limited
resolution of ground-based data.  Attempts to recover core morphology of M31
globular clusters using seeing deconvolution of ground-based data have met only
with a limited success so far (Cohen\&Freeman 1991; Bendinelli \etal 1990).
The HST is obviously a powerful new tool with which to address this problem,
and this has indeed been one of our scientific goals.

So far, we have identified one collapsed-core candidate from among the 13
objects studied, viz., G 105 = Bo 343 = V 199.  Its photometric properties are
rather similar to the other objects in our sample (Tables 4, 6, 7).
Image deconvolutions using all three of our techniques have been performed, and
found to be in an excellent mutual agreement (see Figure 7), especially in the
central regions, where the signal-to-noise ratio is high, and which are
the relevant ones for the studies of core morphology.
We performed tests using a simulated
post-core-collapse cluster, ``moved'' to the distance of M31, convolved with
the observed PSF, and then restored.  All three deconvolution methods mutually
agree, and clearly show the presence of a central power-law cusp, both for the
simulated collapsed-core cluster, and G 105 = Bo 343.  The other clusters in
our sample do not show such core morphology, down to our resolution limits.
\begfig 2 cm
\figure{7} {{\it from left to right}: Observed {\it (squares)} and
RMG-deconvolved {\it (line)} profile of the M31 cluster G 105 = Bo 343;
deconvolved surface brightness profiles of G 105 = Bo 343,
using the three techniques and shifted vertically for clarity
({\it line:} RMG-deconvolved profile, {\it open squares:} MEM,
{\it filled squares:} RL);
the radial distribution of the residuals of
the deconvolved profiles of the cluster G 105 = Bo 343 after  normalization
and calibration (see Sect. 3.6; {\it Open squares:} MEM $-$ RMG;
{\it filled squares:} RL $-$ RMG). }
\endfig
The apparent central density cusp in G 105 = Bo 343 has a power-law slope
$\simeq -$0.75, uncertain by about 10\%, and an unresolved core.  This is a
typical slope for the central cusp profiles of post-core-collapse clusters in
our Galaxy (Djorgovski\&King 1984, 1986; Djorgovski 1988).  Curiously, it is
also the power-law slope characteristic of star clusters around massive black
holes (Bahcall\&Wolf 1976); perhaps deep Rosat X-ray observations of this
cluster might be of some interest.  The cluster is not resolved down to the
limit of our data, i.e., $r_c <$ 0.05 arcsec, or roughly 0.2 pc at the distance
to M31; this is perfectly consistent with the observed core radii or upper
limits on them for the Galactic post-core-collapse clusters.  We note that the
PSF mismatching would in general tend to {\it smear} the deconvolved images and
profiles, and thus hide the central cusp or the unresolved core.  If the PSF
were at fault, we would have probably seen similar effects in our exposures of
other clusters, and we do not.
We also feel fairly confident in excluding the presence of a single star of
V = 20 -- 21 (corresponding to $\sim$ 2\% of the total cluster light), either
a cluster
RG or a field (M31 or galactic) star, projected very close to the cluster
center. Our simulations show that such a star is not able to provide enough
light to transform a King profile into a power-law profile extended over a
few parsecs (i.e. a few tens pixels). This can also be seen by comparing
G105 with G11,
a very similar cluster in all other respects, which is about 2 magnitudes
fainter in the central surface brightness. On the other hand, the presence
of a brighter star would have been easily detected from its spiky--profile
in the deconvolved image, as it was the case of the foreground star in
G11 (see Sect. 3.4).
We thus think that the case for a collapsed core
in G 105 = Bo 343 is quite good, and conclude that this is the first detection
of a collapsed cluster core outside our Galaxy or its nearest satellites.

The significance of this result is that it opens the possibility of
morphological classification of globular clusters in M31, much in the same way
as it was done for our Galaxy using ground-based observations.  This, in
turn, would lead to a better understanding of the globular cluster system of
M31, its dynamical evolution, and comparisons with that of our own Galaxy.  We
hope that more similar cases might turn up as our sample size increases in
the future.

\titleb{Photometric calibrations}
\titlec{Instrumental magnitudes and color transformation}
After removing all the instrumental effects, a direct calibration of
star counts into magnitudes can be obtained by inserting the appropriate
quantities into the FOC $f/96$ calibration equation (Nota 1993,
private communication):
$$
B_{HST} = -21.10 - 2.5 \log [{({(F-BKG)\times U)}/{(\epsilon \times t)}}]
\eqno(1)
$$
\noindent
where $F$ is the total flux (counts) sampled by the fraction of the
star PSF actually considered, $BKG$ is the corresponding background
contribution, $U$ is the inverse sensitivity of the global observational
setup (including the neutral filters) as reported in the header of
the image, $\epsilon$ is the fraction of the total star energy actually
considered, and $t$ is the exposure time.
\par
Due to the HST optical aberration which spreads out the star energy and the
quite small extension of the observed field, the correct measurement of
both $BKG$ and $\epsilon$ is not an obvious task for these two bright stars.
In fact, the stars occupy almost the whole 512$\times$512 pixel field and,
moreover, since they were observed slightly off-center, the images are
not centrally symmetric. To minimize the effects induced on the measure
of the star magnitude by these problems, we have adopted
an iterative procedure of aperture photometry which allows us to
optimize the estimates of $BKG$ and $\epsilon$ to be used in the above
general formula.
\par
The procedure is based on the idea that, if one knows what fraction of
the star light has actually been sampled and if the estimate of the
background has been made correctly, the resulting magnitude of the
measured object is independent of the size of the aperture.
The fraction of encircled light at varying apertures is given by the PSF
plotted in Figure 26 of the FOC Handbook (Version 3.0, April 1992, p.50), and
the background  has been adjusted iteratively until a value was found
which yielded a constant magnitude independent of the aperture
used in the measure.
We have thus derived $B_{HST} = $14.56 and $B_{HST} = $15.20
for GD 248 and G 158-100, respectively (see Table 5).
\begtab 2 cm
\tabcap{5}{Results of the iterative procedure to carry out the
aperture-photometry for the two standard stars.
The head values of columns 3-6 represent four trial
values for the background (counts s$^{-1}$ px$^{-1}$). We have adopted the
values in column 5 for both stars, which produce stable magnitudes
independently of the aperture size reported in column 1.}
\endtab
Since the ground-based magnitudes and colors of these two stars in the Johnson
standard system are known, $B_{J} = $15.20, $B-V = $0.09,
and $B_{J} = $15.58, $B-V = $0.69, respectively,
these two objects can also yield a very rough color equation useful
to transform $HST(F430W)-$magnitudes into the standard ones. In Figure 8
we report the difference $B_{HST} - B_{J}$ for the two observed stars
and for a grid of model stars (F0V-K7V) taken from Schmidt-Kaler
(1982) and convolved with the FOC $f/96 +$ F430W passband (De Marchi 1993,
private communication).
As one can see, the overall trend is fairly consistent and well defined,
indicating the higher sensitivity of the HST photometric system in the blue
range, as expected. The resulting color equation
$$
\Delta B_{(HST-J)} \sim 0.64 (B-V)_0 - 0.80  \eqno(2)
$$
\noindent
represents a preliminary result and needs more work for further confirmation,
however it is quite adequate to our purpose of transforming the magnitudes
between the two photometric systems.
\begfig 2 cm
\figure{8~} {Adopted color transformation between the HST and ground-based
standard systems. {\it Dots:} grid of model stars (F0V-K7V)
convolved with the FOC f/96 $+$ F430W; {\it Stars:} GD 248 and G158-100,
calibration stars for the PSF. The dashed line is the best fit to the model
stars, given as Eq. 2 in the text.}
\endfig
\titlec{Integrated magnitudes of globular clusters}
\medskip\noindent
The procedure described in the previous section cannot be applied identically
to the globular clusters, because they are extended objects and tend
to fill the image, thus making the estimate of $BKG$ and $\epsilon$ rather
inaccurate.

For the globular clusters we have therefore estimated the background and
the fraction of sampled energy from the analysis of the observed profiles and
then we have derived the $B_{HST}$ integrated magnitudes via Eq. (1).
These quantities are listed in Table 6 in columns 3 and 4.
\begtab 2 cm
\tabcap{6}{HST-integrated magnitudes of M31 globular clusters.}
\endtab
Given the uncertainties affecting the estimate of $BKG$ and $\epsilon$,
these magnitudes are less accurate than those obtained for the
standard stars. We estimate that the errors on $BKG$ and $\epsilon$ are
$\le$ 10\%, leading to an error of about 10\% ($\sim$ 0.1 mag) in the
integrated flux (magnitude) determination.

A check can be performed using the $(B-V)$ colours available from the
ground-based photometry and the colour equation derived in the previous
section to determine the HST-magnitudes transformed to the Johnson
system, $B_{cor}$.
The final error on the $B_{cor}$ magnitudes thus derived is slightly
larger than for the corresponding B$_{HST}$ values, due to the additional
errors on the colours and on the definition of the colour equation.
These $B_{cor}$ magnitudes are listed in column 5 of Table 6 and compare
very well with the ground-based values reported in column 6,
with the only exception of Bo 289.
For this cluster a significant amount of light should have been
lost in the HST frame (approximately 25\%  in order to reconcile the
HST and ground-based magnitudes), which is rather unlikely and
does not appear from the inspection of its profile.
Alternatively, the ground-based $B$ magnitude could have been underestimated
by about 0.25 mag, which is not impossible since the only available measure
is based on photographic photometry (Battistini \etal 1987), and a redder
colour would be more consistent with the estimated metal abundance
(Bonoli \etal 1987).

\titlec{Systematic biases in HST surface photometry using deconvolutions}

\medskip\noindent
While deconvolution techniques are used to restore the {\it shapes} of
surface brightness profiles, they do not guarantee photometric, e.g.,
magnitude zero-point accuracy.  In fact, in surface photometry applications
such as ours, a {\it systematic bias} may be present, as briefly explained
below.  We intend to address this question in more detail in a forthcoming
publication.

Methods such as RL or MEM are typically programmed in such a way so that the
total signal in the output (deconvolved) image is the same as in the input
image.  For many applications this is indeed appropriate.  However, if a
non-negligible fraction of the light from the object(s) imaged spills outside
the image boundaries, either due to the object extent and/or the large PSF
wings (both of which are the case here), a bias will result. Deconvolution
techniques cannot account for the signal which has been removed outside the
image boundaries.  For an image where an extended, bright object is present,
the amount of light moved outside the image boundaries by the convolution
exceeds the amount of light brought in. Thus, there is a net deficiency of
the integrated signal inside such an image, which cannot be recovered by the
deconvolution techniques (since they ``do not know'' about the outside).  Thus,
the net effect would be to {\it lower} the deconvolved profile, in the amount
roughly proportional to the amount of light scattered outside the image
boundaries, while giving it more or less the correct shape.  The effect should
be more pronounced near the image edges, where the relative loss of light would
be higher, and the shape of the profile would be affected as well.  This bias
is an inevitable artifact of the demand that the input and the output images
have the same flux; if the image contains object(s) brighter than their
surroundings, the deconvolved image should have a higher total flux. For images
of relatively uniform surface brightness and star density, or with very compact
PSF's, the net effect should be negligible.

Interestingly enough, this bias is expected to be less important in a technique
such as the RMG method, where surface brightness profile is implicitly
extrapolated to infinity, albeit with a Gaussian falloff.  In a sense, some
fraction of the missing light is accounted for by the assumed functional form
of the profile. Some preliminary tests have been performed using a simulated
image of an extended object, which was deconvolved by the three deconvolution
methods on subimages of size 512x512 and 256x256 pixels.  Although the tests
are presently rather crude and must be taken only as a general indication (more
accurate ones are in progress and will be presented in a forthcoming paper),
they do show this {\it edge effect} which becomes more pronounced as the array
gets smaller, and is strongest in the RL and MEM methods.  The RMG
deconvolutions seem to lead to a negligible bias of this type. A similar
consistent indication can be derived by the inspection of the radial behaviour
of the residuals (RL $-$ RMG, MEM $-$ RMG) for the individual clusters shown in
Figure 6. In fact, as can be seen in all panels, the residuals tend to diverge
at the border of the images. On this basis, we will use the RMG results as
fiducial for the purpose of defining the profile intensity zero-points.

\titlec{Central surface brightness from calibrated deconvolved profiles}

Following the arguments given in the previous section, we adopt as a reference
profile for all the clusters the RMG deconvolved profile extending out to
$r_{lim} \sim$ 1.5 arcsec.  At these radii all the deconvolved profiles are in
a good mutual agreement (see Figure 6), and are relatively unaffected by the
edge effects described above.  If the integrated flux within the $r_{lim}$
aperture were not the same for all deconvolution methods, then a normalization
to the RMG value would be necessary.

The absolute calibration of the deconvolved profiles can thus be obtained using
the relation:
$$
\mu_{0} = B_{STAR} - 2.5 \log \left[ {C_{0} \over C_{STAR}} \times
\left( {L \over L_{RMG}} \right) ^{-1} \right] ~~,
\eqno(3)
$$
\noindent
where $\mu_0$ is the cluster central surface brightness in $B_{HST}$ magnitudes
per square arcsec, $C_0$ the peak pixel intensity in counts/sec as estimated
from the central pixel, and $B_{STAR}$ and $C_{STAR}$ the corresponding
magnitude and total counts/sec in the $B_{HST}$ band for a calibration star.
We have used GD 248 for this calibration; the other standard star gives the
same results within 0.01 mag.  Finally, ${L \over L_{RMG}}$ is the
normalization factor, i.e., the ratio of the integrated flux in counts/sec
encircled within $r_{lim}$ as obtained from the RL and MEM profiles with
respect to the RMG profile.  The central surface brightness values $\mu_{B}(0)$
(in the HST photometric system) derived in this way are listed in Table 7.
\begtab 2 cm
\tabcap{7}{Intrinsic parameters of the deconvolved M31 clusters (1 pc =
0.294 arcsec, corresponding to a distance of 700 kpc).}
\endtab
Table 7 lists the core radius $r_c$, the central surface brightness $\mu_0$,
and the radius $r_h$ containing a half of the integrated projected light for
the clusters (see Sect. 3.6.2), assuming a distance of 700 kpc to M31.  Figure
9 shows the comparison of the values obtained for the same clusters from the
different procedures.  The overall correspondence between different methods
is quite satisfactory.
\begfig 2 cm
\figure{9~} {Comparisons of the results obtained for the cluster
intrinsic parameters using the three  deconvolution
methods. {\it Top:} Central surface brightness; {\it middle:}
half-light radius; {\it bottom:} core radius.}
\endfig

In Table 8 we list the measurements of the half width at half maximum (HWHM,
or effectively the core radius) from the raw profiles, and the three
deconvolution methods.  The overall agreement between the different methods is
quite good for most clusters.  The differences in core radii between the
different deconvolution methods are of the order of 20\%, and this is probably
indicative of the measurement errors.  We also note that the deconvolved data
have core radii typically a factor of two smaller than the raw data, even at
the present HST resolution!  This confirms the necessity of deconvolutions in
this work.
\begtab 2 cm
\tabcap{8}{Comparison between HWHM (namely the core radius $r_c$)
of the observed and deconvolved profiles.}
\endtab
The resulting values of $\mu_0$ derived using RL method agree with the results
of the RMG method on the average within 0.16 $\pm$ 0.09 mag, and the values
derived using MEM to within 0.06 $\pm$ 0.15 mag (the quoted uncertainties
indicating the spread of the $\Delta\mu$ values).
The zero-point shifts are actually controlled by the normalization procedure
and, in particular, by the adopted radial extension of the profile, $r_{\lim}$.
For instance, if the normalization is extended out to the image borders
(including thus also the quoted {\it edge effects}), the
MEM (and to a lesser degree also the RL) deconvolutions tend to produce
slightly brighter values of central surface brightness than the RMG method.
These methods are generally optimised to ``pull out'' point sources, and so may
be slightly biased in favour of high spatial frequencies, yielding sharper
image
cores.
\titlea {Structural parameters of clusters as derived from surface photometry}

\titleb{The intrinsic parameters of the deconvolved clusters}

The main goal of this project so far was to measure structural parameters of
globular clusters in M31, to compare them with those of Galactic globulars, and
explore any possible correlations which may be found.  Given the limited sample
of only 13 objects, we can only make a preliminary inquiry of this type.

There are basically three independent quantities which we can measure from our
data: the core radius, $r_c$, operationally defined as the HWHM of the
deconvolved profiles, the half-light radius, $r_h$, and the projected central
surface brightness, $\mu_0$.  They are listed in Table 7.  Mean surface
brightness within the half-light radius, $\langle \mu \rangle _h$ is also
measured (see Table 10),
but given that clusters are presumably well described by the
3-parameter family of King (1966) models, this quantity is not really
independent of the others.  To them we can also add the total absolute
magnitudes from the ground-based measurements, $M_V$, listed in Table 1 (the
same caveat applies).  In principle, the ratios of the half-light and core
radii can also be used to estimate the King concentration parameters, $c$, but
obviously this brings no additional independent information.

\titleb{Comparisons with the ground-based studies of M31 globulars}

The only study to date of which we are aware of, in which seeing deconvolution
is used to measure the structural parameters of M31 globular clusters is that
by Bendinelli \etal (1990).  Unfortunately, we have no targets in common.  We
note however, that our derived structural parameters are of a similar
magnitude.

Six previous studies (Battistini \etal 1982, hereafter B82; Crampton \etal
1985, hereafter C85; Spassova \etal 1988; Lupton 1989; Davoust\&Prugniel
1990, hereafter DP90; and Cohen\&Freeman 1991) presented estimates of the
structural parameters of several M31 clusters, usually derived using some
indirect method, and often assuming that King (1966) models are a good
descriptor of the cluster profile morphology (this is certainly a good
approximation for the Galactic globulars).  In Table 9 we list the data for
the clusters we have in common with these studies.
\begtab 2 cm
\tabcap{9}{Comparison with previous ground-based data for the common
clusters.}
\endtab
If we look at the ground-based results alone, we see no significant systematic
differences between the $r_c$ values obtained for the clusters in common by B82
and Cohen\&Freeman (1991).  Those obtained by C85 are systematically larger
(about 30-50\%), while DP90 report ``corrected'' values which are on the
average 0.2-0.3 pc smaller.

As far as our HST results are concerned, we can compare our core radii only
with those estimated by B82, C85 and DP90.  We have 5 clusters in common with
both B82 and C85, i.e. Bo 6 = G 58, Bo 12 = G 64, Bo 23 = G 78, Bo 193 = G 244,
and Bo 218 = G 272, plus 2 additional clusters in common with C85, i.e., Bo 373
= G 305, and Bo 386 = G 322.  The clusters Bo 12 and Bo 23 are also in common
with DP90. For them, the quoted core radii are generally larger than our
deconvolved HST measurements, as it may be expected.  All that we can conclude
at this point is that uncompensated ground-based measurements are likely to
overestimate the core radii by at least 30\%, and generally more. This bias is
{\it systematic}, and would strongly affect any relevant comparisons of the
Galactic and M31 globular cluster systems.

\titleb{Comparison with the properties of Galactic globulars}

It is of some interest to compare the properties we have derived for the M31
globulars with those of their Galactic counterparts.  For this purpose we use a
recent data compilation on Galactic globulars by Djorgovski (1993).  We have
converted our $B$-band HST measurements to the $V$-band (in which the data for
Galactic clusters are best defined) using the global $(B-V)$ colors of M31
globulars as measured from the ground, which are given in Table 1.  In Figure
10 we compare the distributions of global properties of our M31 clusters
measured so far with those of the Galactic globular clusters observed from the
ground.
\begfig 2 cm
\figure{10~} {A comparison of the global properties of Galactic globular
clusters (crosses) with the measurements for our sample of M31 clusters
(solid squares).  We have used the RMG deconvolution results, and converted
to the $V$ band using global, ground-based $(B-V)$ colors.  Measurements
of the core radius, $r_c$, half-light radius, $r_h$, central surface
brightness, $\mu_V(0)$, and average surface brightness within $r_h$,
$\langle \mu \rangle _{hV}$, are plotted versus the cluster absolute
magnitude, M$_V$, in panels (a) -- (d), respectively.  Correlation between
the core parameters, $r_c$ and $\mu_V(0)$, is shown in panel (e), and the
correlation between the two characteristic radii in panel (f).  Note that
our sample of M31 globulars occupies the region covered by the brighter
Galactic globulars, without any apparent systematic differences.}
\endfig
\begtab 2 cm
\tabcap{10}{Cluster data and parameters from RMG deconvolutions.}
\endtab
We see that the M31 globulars from our sample occupy a portion of the parameter
space corresponding to the brighter Galactic clusters, which is simply the
sample selection effect.  Our sample clusters have core and half-light radii,
and central and half-light average surface brightness values perfectly normal
for their luminosities (Fig. 10abcd).  Their core parameters seem to be
correlated in the same way as those of the Galactic globulars (Fig. 10e).  They
also have normal half-light radii for their core sizes (Fig. 10f), albeit with
a smaller spread; this may be an accident, or a selection effect, or even
partly an artifact of our observing and data reduction process (e.g., due to a
finite and small detector size).

We cannot say much about the correlations between different cluster properties,
because of a small dynamical range of cluster luminosities in our sample.  They
certainly seem to follow the overall trends set by the Galactic globulars, but
in order to test this, we would need to extend our observations to some fainter
clusters.  This is essentially $the$ limiting factor in this study so far.

The fact that M31 globulars seem to follow the overall distributions of, and
correlations between various global properties of Galactic globulars is
reassuring, and gives us an added confidence in our measurements and
deconvolution procedures.  We are at least beginning to study the physical and
morphological properties of M31 globulars, with an accuracy which is clearly
adequate for the achievement of our scientific goals.

\titlea{Conclusions and future prospects}

Our basic conclusion is that excellent surface photometry of M31 globular
clusters is feasible using HST data, even with the present problems with its
imaging capability.  Three independent deconvolution techniques used here lead
to perfectly consistent results.  This opens the possibility of morphological
studies of M31 globular clusters comparable to the ground-based studies of
Galactic globulars.  We have also demonstrated that a failure to perform
accurate and adequate seeing deconvolutions in ground-based studies of M31
globulars can lead to systematically biased measurements of their structural
parameters.  On the other hand, quantitative measurements of cluster structural
and photometric parameters, and even detection of the characteristic
post-core-collapse central cusps (if present) can be reliably made using HST
observations.

Whereas the sample studied here is rather small, our preliminary comparisons
with the global properties of Galactic globular clusters show that the two
systems consist of very similar objects.  So far, the similarity of globular
cluster systems in different galaxies was largely confined to their luminosity
functions and metallicity distributions (cf. Harris 1991, and references
therein).  Here we compared for the first time, albeit with a limited sample,
the distributions of cluster structural parameters (characteristic radii and
surface brightness values) as well.

As far as future developments are concerned,
the feasibility of an adequate stellar photometry in these data remains
unclear.  Whereas it is clearly possible, with the aid of image deconvolution
techniques, to detect and measure at least some individual stars in these
clusters, many issues need to be explored further.  This will include a careful
analysis of the noise properties, completeness and false detections, the
reliability and possible biases in the stellar photometry on the deconvolved
images, etc.  More extensive simulations are needed before any firm conclusions
or quantitative results can be presented, and efforts in this direction
are now in progress.

It is thus possible to conduct fruitful surface photometry and morphology
studies, and maybe also stellar photometry studies of M31 globulars with the
HST even in its present state, with a level of accuracy and reliability
surpassing that of the ground-based studies.  If the HST repair mission is
successful, the situation should be considerably improved.  We then hope to
resume our original program with the three basic scientific goals as described
in the introduction, especially the CMD studies below the HB magnitude level.

\bigskip\noindent
\bigskip\noindent
\ack{It is a pleasure to thank I.R. King and  R. Walterbos
for the work and efforts in the early phases of this project, and the
continuous support and useful discussions since.  We are also deeply
grateful to A. Nota, G. De Marchi, and A. Pasquali for help at various stages
of this work.
This research was supported
by the Consiglio delle Ricerche Astronomiche (CRA) of the Ministero delle
Universit\'a e della Ricerca Scientifica e Tecnologica (MURST) in Italy.
Additional support for this work was provided by NASA through grant number
GO-2583 from the Space Telescope Science Institute, which is operated by the
Association of Universities for Research in Astronomy, Inc., under NASA
contract NAS5-26555.}

\begref
\ref Bahcall J.N., Wolf R.A., 1976, ApJ 209, 214
\ref Battistini P., B\`onoli F., Braccesi A., Fusi Pecci F., Marano B.,
1980, A\&AS 42, 357
\ref Battistini P., B\`onoli F., Buonanno R., Corsi C.E., Fusi Pecci F.,
1982, A\&A 113, 39
\ref Battistini P., B\`onoli F., Braccesi A., Federici L., Fusi Pecci F.,
Marano B., Borngen F., 1987, A\&AS 67, 447
\ref Battistini P., B\`onoli F., Casavecchia M., Ciotti L.,
Federici L., Fusi Pecci F., 1993a, A\&A 272, 77
\ref Battistini P. \etal, 1993b, in preparation
\ref Bendinelli O., 1991, ApJ 366, 599
\ref Bendinelli O., Cacciari C., Djorgovski S., Federici L., Ferraro
F., Fusi Pecci F., Parmeggiani G., Weir N., Zavatti F., 1992, in:
Benvenuti P., Schreier E. (eds.) Proc. ST-ECF/STScI Workshop {\sl Science
with the Hubble Space Telescope}, ESA publication, p.271
\ref Bendinelli O., Cacciari C., Djorgovski S., Federici L., Ferraro
F., Fusi Pecci F., Parmeggiani G., Weir N., Zavatti F., 1993,
ApJ 409, L17
\ref Bendinelli O., Parmeggiani G., Zavatti F., Djorgovski S., 1990,
AJ 99, 774
\ref B\`onoli F., Delpino F.E., Federici L., Fusi Pecci F., 1987, A\&A
185, 25
\ref Brodie J.P., 1993, in: Smith G., Brodie J.P. (eds.) The Globular
Cluster $-$ Galaxy Connection, ASPCS, p.483
\ref Buonanno R., Buscema G., Corsi C.E., Ferraro I., Iannicola G.,
1983, A\&A 126, 278
\ref Buonanno R., Corsi C.E., Battistini P., B\`onoli F., Fusi Pecci F.,
1982, A\&AS 47, 451
\ref Buonanno R., Iannicola G., 1988, PASP 101, 294
\ref Cacciari C., Battistini P., Bendinelli O., Bonoli F., Buonanno R.,
Djorgovski S., Federici L., Ferraro F., Fusi Pecci F., King I.R.,
Parmeggiani G., Walterbos R., Zavatti F., 1992, in:
Benvenuti P., Schreier E. (eds.) Proc. ST-ECF/STScI Workshop {\sl Science
with the Hubble Space Telescope}, ESA publication, p.157
\ref Chernoff D., Djorgovski S., 1989, ApJ 339, 904
\ref Cohen J.G., 1993, in:  Smith G., Brodie J.P. (eds.) The Globular
Cluster $-$ Galaxy Connection, ASPCS, p.438
\ref Cohen J.G., Freeman K.C., 1991, AJ 101, 483
\ref Collura A., Reale F., Peres G., 1990, ApJ 154, 891
\ref Crampton D., Cowley A.P., Hutchings J.B., Schade D.J., Van Speybroeck
L.P., 1984, ApJ 284, 663
\ref Crampton D., Schade D.J., Chayer P., Cowley, A.P., 1985, ApJ 288, 494
\ref Christian C.A., Heasley J.N., 1991, AJ 101, 848
\ref Davoust E., Prugniel P., 1990, A\&A 230, 67
\ref Djorgovski S., 1985, Ph.D. thesis, Univ. of California, Berkeley
\ref Djorgovski S., 1988, in:
Grindlay J.E., Philip S.G.D. (eds.)IAU Symp. No. 126,  Globular
Cluster Systems in Galaxies, Dordrecht, Kluwer, p.333
\ref Djorgovski S., 1993, in: Djorgovski S., Meylan G. (eds.) Structure
and Dynamics of Globular Clusters, ASPCS, 50, 373
\ref Djorgovski S., King I.R., 1984, ApJ 277, L49
\ref Djorgovski S., King I.R., 1986, ApJ 305, L61
\ref Djorgovski S., Piotto G., Phinney E.S., Chernoff D.F., 1991, ApJ
372, L41
\ref Djorgovski S., Piotto G., 1992, AJ 104, 2112
\ref Djorgovski S., Piotto G., 1993, in: Djorgovski S., Meylan G. (eds.)
Structure and Dynamics of Globular Clusters, ASPCS, 50, 203
\ref Elson R., Hut P., Inagaki S., 1987, ARA\&A 25, 565
\ref Federici L., Fusi Pecci F., Marano B., 1990, A\&A 236, 99
\ref Federici L., B\`onoli F., Ciotti L., Fusi Pecci F., Marano B.,
Lipovetsky V.A., Neizvestny S.L., Spassova N., 1993, A\&A 274, 87
\ref Fusi Pecci F., Cacciari C., Federici L., Pasquali A., 1993a, in:
Smith G., Brodie J.P. (eds.) The Globular
Cluster $-$ Galaxy Connection, ASPCS, p.410
\ref Fusi Pecci F., Ferraro F.R., Bellazzini M., Djorgovski S.,
Piotto G., Buonanno R., 1993b, AJ 105, 1145
\ref Harris W.E., 1991, ARAA 29, 543
\ref Heasley J.N, Christian C.A., Friel E.D., Janes K.A., 1988,
AJ 96, 1312
\ref Huchra J.P., 1993, in:
Smith G., Brodie J.P. (eds.) The Globular
Cluster $-$ Galaxy Connection, ASPCS, p.420
\ref Huchra J.P., Brodie J.P., Kent S.M., 1991, ApJ 370, 495
\ref King I.R., 1966, AJ 71, 64
\ref King I.R., Stanford S.A., Seitzer P., Bershady M., Keel W., Koo D.,
Weir N., Djorgovski S., Windhorst R., 1991, AJ 102, 1553.
\ref Long K.S., van Speybroeck L.P., 1983, in: Lewin W., van den Heuvel
E.P.J. (eds.) Accretion Driven X-ray Sources, Cambridge University
Press, p. 41
\ref Lucy L. B., 1974, AJ 79, 745
\ref Lucy L. B., 1992, AJ 104, 1260
\ref Lupton R.H., 1989, AJ 97, 1350
\ref Meylan G., Djorgovski S., 1987, ApJ 322, L91
\ref Monnet G., Bacon R., Emsellem E., 1992, A\&A 253, 366
\ref Peterson C., Reed C., 1987, PASP 99, 20
\ref Racine R., 1993, in: Smith G., Brodie J.P. (eds.) The Globular
Cluster $-$ Galaxy Connection, ASPCS, in press
\ref Racine R., Harris W.E., 1991, CFHT Inf. Bull. No. 24, p.17
\ref Richardson W.H., 1972, JOSA 62, 55
\ref Sargent W.L.W., Kowal C.T., Hartwick F.D.A., van den Bergh S.
1977, AJ 82, 947
\ref Schmidt-Kaler T.H., 1982, in: Schaifers K., Voigt H.H. (eds.)
Landolt-Bornstein New Series, Vol. 2b, Astronomy and
Astrophysics -- Stars and Star Clusters, New York, Springer-Verlag
\ref Spassova N.M., Staneva A.V., Baev P.V., 1988, in:
Grindlay J.E., Philip S.G.D. (eds.) IAU Symp. No. 126, Globular
Cluster Systems in Galaxies, Dordrecht, Kluwer, p.569
\ref Thomas P., 1989, MNRAS 238, 1319
\ref Trinchieri G., Fabbiano G., 1991, ApJ 382, 82
\ref van den Bergh S., Morbey C., Pazder J., 1991, ApJ 375, 594
\ref van Speybroeck L.P., Epstein A., Forman W., Giacconi R.,
Jones C., Liller W., Smarr L., 1979, ApJ 234, L45
\ref Weir N., 1991, in: Grosbol P., Warmels R. (eds.) Proc. of the 3rd
ESO/ST-ECF Data Analysis Workshop. Garching, ESO, p.~115.
\ref Weir N., Djorgovski S., 1990, in: White R.L., Allen R.J. (eds) Proc. of
a STScI Workshop, The Restoration of HST Images and Spectra,
Baltimore, STScI, p.~31.
\ref Zavatti F., Bendinelli O., Gatti M., Parmeggiani G., 1991, in:
Grosbol P., Warmels R. (eds.) Proc. of the 3rd ESO/ST-ECF Data Analysis
Workshop. Garching, ESO, p.~179.
\endref
\end
\bye